\documentclass[aps,prl,twocolumn,tightlines]{revtex4}%
\usepackage{amsfonts}
\usepackage{amsmath}
\usepackage{amssymb}
\usepackage{graphicx}%
\begin{document}
\title{Beyond the entanglement of qubit pair in a mixed state}
\author{S. Shelly Sharma$^{\dag}$}
\affiliation{Departamento de F\'{\i}sica, Universidade Estadual de Londrina, Londrina
86051-990, PR Brazil}
\author{N. K. Sharma$^{\ddag}$ }
\affiliation{Departamento de Matematica, Universidade Estadual de Londrina, Londrina
86051-990, PR Brazil }

\begin{abstract}
Given a multipartite quantum system that consists of two-level particles
(qubits), one may or may not have access to all the subsystems. What can we
know about the entanglement of the multiqubit system and residual correlations
beyond two-tangle if we have access only to two-qubits at a time? Algebraic
analysis of two-qubit states yields monogamy constraints on distribution of
entanglement between sub-systems of an N-qubit state and criterian to
determine if the state has multipartite entanglement. Monogamy constraints,
reported in this letter, are relations between well known entanglement
measures such as one-tangle, two-tangles and three-tangles of an N-qubit pure state.

\end{abstract}
\maketitle


\paragraph*{\textbf{\textit{Introduction:}}}

Quantum entanglement was pointed out by Schr\"{o}dinger \cite{schr35} to be a
characteristic feature of quantum systems. During the last three decades, it
has come to be viewed as an essential physical resource in technologies like
quantum computations \cite{shor94} and quantum cryptography \cite{benn84} with
known advantages over their classical counterparts. Quantum entanglement also
has important applications in other areas such as quantum field theory
\cite{cala12}, statistical physics \cite{sahl15}, and quantum biology
\cite{lamb13}. Bipartite entanglement is well understood as there is concise
result to entanglement classification problem. However, the characterization
of multipartite entanglement which is a resource for multiuser quantum
information tasks \cite{xue02, xue17,yang20,louk10,ma08}, is a far more
challenging task \cite{horo09}.

A physical qubit is any two-level quantum system with states represented by
kets $\left\vert 0\right\rangle $ and $\left\vert 1\right\rangle $. Given a
multipartite systems that consists of qubits, one may or may not have access
to all the subsystems. In an insightful article Walter et al. \cite{walt13}
used an algebraic geometry approach to show that states of individual
particles are a rich source of information on multiparticle entanglement. What
can we learn about the multipartite entanglement if we have access to only two
qubits at a time? In this letter, properties of two-qubit states are shown to
yield constraints on entanglement measures (tangles) of an $N-$qubit pure
state. For a selected focus qubit, there are $N-1$ possible two-qubit states.
Total entanglement of the focus qubit with rest of the system is quantified by
one-tangle. For example if a qubit in state $\rho_{1}$ at location $A_{1}$ is
the focus qubit then one tangle is $\tau_{1|2..N}=4\det\rho_{1}$. The
entanglement of two qubits at $A_{1}$ and $A_{j}$ in state $\rho_{1j}$ is
known to be quantified by two-tangle $\tau_{1|j}$ (or concurrence as defined
in ref. \cite{hill97,woot98} ). Three-tangle $\tau_{i|j|k}$ \cite{coff00} is a
measure of three-way entanglement of qubits. Entanglement due to four-way
correlations may likewise be quantified by four-tangles and so on so forth.

The value of $\tau_{1|j}$ is determined by eigenvalues of matrix $\rho
_{1j}\widetilde{\rho_{1j}}$ , where $\widetilde{\rho_{1j}}=\left(  \sigma
_{y}\otimes\sigma_{y}\right)  \rho_{1j}^{\ast}\left(  \sigma_{y}\otimes
\sigma_{y}\right)  $. Here the action of Pauli operator $\sigma_{y}%
$=$i\left\vert 1\right\rangle \left\langle 0\right\vert -i\left\vert
0\right\rangle \left\langle 1\right\vert $ on a qubit is to swap its state and
$\ast$ stands for complex conjugation. Coefficients in the characteristic
polynomial of $\rho_{1j}\widetilde{\rho_{1j}}$ depend on the mixedness of
state $\rho_{1j}$ which is, in turn, a function of entanglement of the pair of
qubits with the rest of the system. Characteristic polynomial of matrix
$\rho_{1j}\widetilde{\rho_{1j}}$ has at the most four polynomial coefficients
$n_{d}\left(  \rho_{1j}\right)  $ (degree $d=4,8,12,$ and $16$). Each term in
$n_{d}\left(  \rho_{1j}\right)  $ is a product of $d$ number of state
coefficients of $N-$qubit pure state. The polynomial coefficients are unitary
invariant functions of state coefficients as such are, by themselves, possible
candidates to characterize the state. However, by writing $n_{d}\left(
\rho_{1j}\right)  $ in terms of state coefficients it is possible to identify
entanglement measures to quantify two-way, three-way,..., N-way entanglement.
A two-qubit state can be obtained from an $N-$qubit pure state in so many
different ways. For example if $N=4$ then there are three different paths to
reach $\rho_{12}$ as shown in Fig. (1). Intermediate subsystems and their
tangles are also shown.

\begin{figure}[th]
\centering
\includegraphics[width=3.0in,height=3.0in]{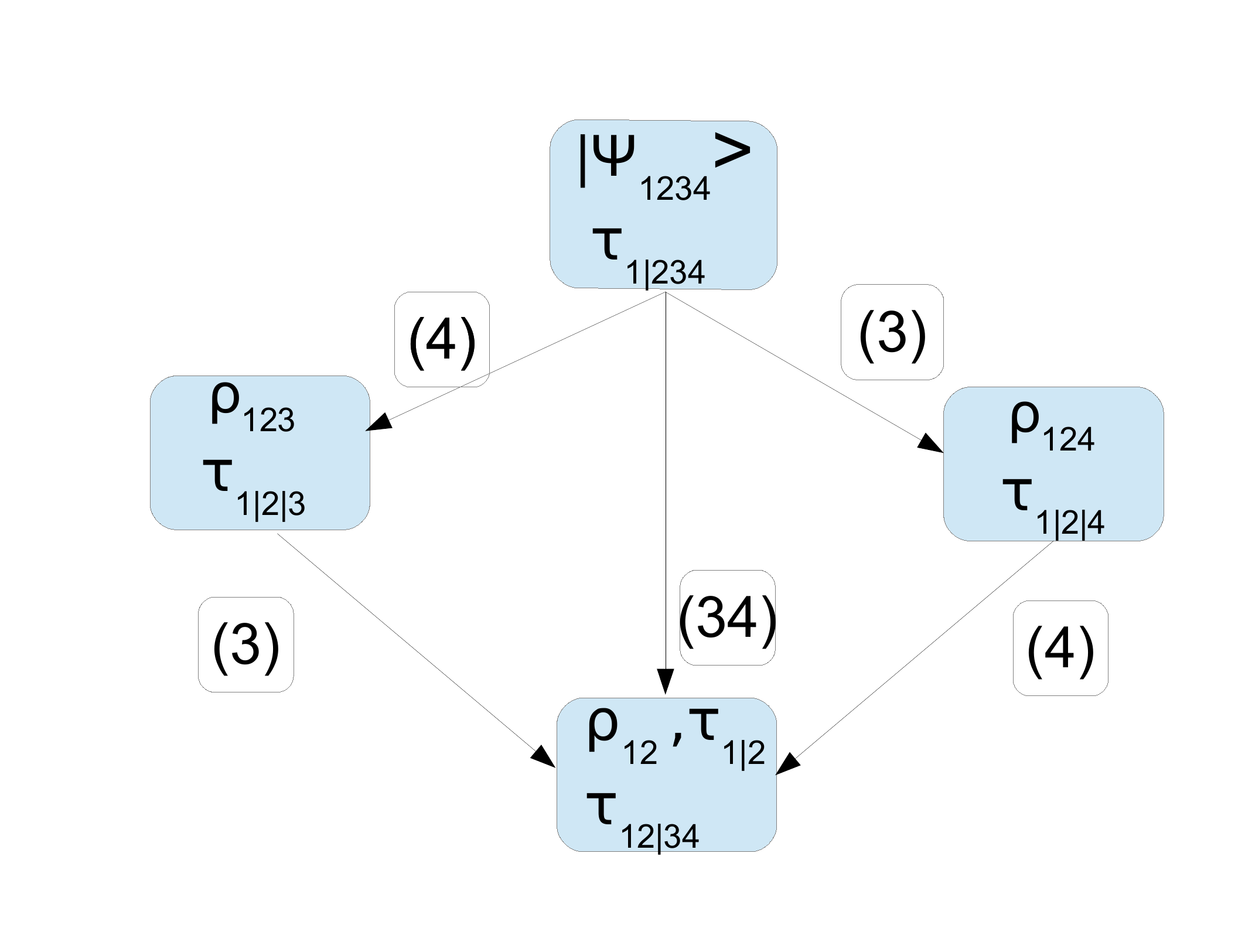}\caption{Paths to
obtain $\rho_{12}$ from four-qubit pure states. The number in bracket refers
to the qubit that has been traced out of $\left\vert \Psi_{1234}\right\rangle
$ }%
\end{figure}

An algebraic analysis of coefficients $n_{d}\left(  \rho\right)  $ brings out
the relation between the tangles of subsystems on all possible paths that lead
from an $N-$qubit pure state to a two-qubit state $\rho$. The main findings of
such an analysis reported in this letter are listed below:

\begin{itemize}
\item[a)] The coefficient $n_{4}\left(  \rho\right)  $ is equal to the sum of
two-tangle and entanglement of the pair with the rest of the system.

\item[b)] The coefficient $n_{8}\left(  \rho\right)  $ contains contributions
from three-way correlations and correlations beyond three-tangles.
Three-tangles of intermediate sub-systems are identified in the expression for
$n_{8}\left(  \rho\right)  $ when composite system has $N\geq3$. Obviously,
$n_{8}\left(  \rho\right)  =0$, if $N-$qubit pure state is separable or does
not contain correlations beyond two-way correlations.

\item[c)] The coefficients $n_{12}\left(  \rho\right)  $ and $n_{16}\left(
\rho\right)  $ contain additional contributions from correlations beyond three-tangles.
\end{itemize}

Combining the results of items a), b) and c) a monogamy constraint on
one-tangle, two-tangles and three-tangles of an $N-$qubit pure state is
obtained. For $N=3$ it reduces to well known monogamy relation for a three
qubit state that is CKW inequality reported in a seminal paper by Coffman,
Kundu, and Wootters \cite{coff00}. For $N>3$, our result extends the monogamy
relation of Osborne and Verstraete \cite{osbo06} and identifies residual
correlations beyond three tangles for $N-$qubit states. As shown in ref.
\cite{arXiv2021v3}, for $N=4$ it is possible to identify distinct four-tangles
in $n_{8}\left(  \rho\right)  $ to quantify four-way correlations in different
classes of four-qubit pure states. Some of the recent efforts on monogamy of
entanglement, where CKW inequality is generalized and also replaced by other
entanglement measures include \cite{regu14,regu16,karm16,shar18,park18,chri18}%
. Monogamy of entanglement has potential applications in areas of physics such
as quantum key distribution \cite{terh04,pawl10,gisi02}, classification of
quantum states \cite{dur00,gior11,prab12}, frustrated spin systems
\cite{ma11,rao13}, and even black-hole physics \cite{lloy14}. Results reported
in this letter, contribute to understanding the entanglement distribution in
composite quantum systems with focus on information obtained from two-qubit subsystems.


\paragraph*{\textbf{\textit{Two-tangle of a pair of qubits:}}}

Two-tangle or concurrence, a well known calculable measure of two-qubit
entanglement \cite{hill97,woot98}, is an entanglement monotone. Entanglement
of qubit $A_{1}$ with $A_{2}$ in a two-qubit pure state $\left\vert \Psi
_{12}\right\rangle =\sum_{i_{1},i_{2}}a_{i_{1}i_{2}}\left\vert i_{1}%
i_{2}\right\rangle $, $i_{m}=0,1$, is quantified by two-tangle defined as
\[
\tau_{1|2}\left(  \left\vert \Psi_{12}\right\rangle \right)  =2\left\vert
a_{00}a_{11}-a_{10}a_{01}\right\vert .
\]
Two-tangle of a mixed state $\rho=\sum\limits_{i}p_{i}\left\vert \phi
_{12}^{\left(  i\right)  }\right\rangle \left\langle \phi_{12}^{\left(
i\right)  }\right\vert $ is constructed through convex roof extension.
Defining $C\left(  \rho\right)  =\sqrt{\lambda_{1}}-\sqrt{\lambda_{2}}%
-\sqrt{\lambda_{3}}-\sqrt{\lambda_{4}}$, two-tangle \cite{hill97,woot98} of a
two-qubit state $\rho_{12}$ is given by%
\begin{equation}
\tau_{1|2}\left(  \rho\right)  =\max\left(  0,C\left(  \rho\right)  \right)  ,
\label{2tangle}%
\end{equation}
where $\lambda_{1}\geq\lambda_{2}\geq\lambda_{3}\geq\lambda_{4}$ are the
eigenvalues of matrix $\rho\widetilde{\rho}$ with $\widetilde{\rho}$ $=\left(
\sigma_{y}\otimes\sigma_{y}\right)  \rho^{\ast}\left(  \sigma_{y}\otimes
\sigma_{y}\right)  $. Here $\ast$ denotes complex conjugation in the standard
basis and $\sigma_{y}$ is the Pauli matrix. In the most general case, the
characteristic polynomial of $\rho\widetilde{\rho}$ satisfies%
\begin{equation}
x^{4}-n_{4}x^{3}+n_{8}x^{2}-n_{12}x+n_{16}=0 \label{poly}%
\end{equation}
where the coefficients $n_{d}$ are known functions of eigenvalues of
$\rho\widetilde{\rho}$.

Matrix elements of a two-qubit mixed state $\rho$ are degree-two functions of
state coefficients of the pure state from which $\rho$ has been obtained. A
given coefficient $n_{d}$ is, in turn, a unitary invariant function of state
coefficients of the pure state of which $\rho$ is a part. We can verify that
for $C\left(  \rho\right)  =\pm\left\vert C\left(  \rho\right)  \right\vert $
the coefficient $n_{4}$ satisfies the relation%
\begin{equation}
n_{4}=\left\vert C\left(  \rho\right)  \right\vert ^{2}+2\sqrt{n_{8}%
+2\sqrt{n_{16}}\pm2\sqrt{f_{16}}}, \label{n4rho}%
\end{equation}
where $f_{16}\geq0$ is given by%
\begin{align}
f_{16}  &  =\sqrt{n_{16}}\left\vert C\left(  \rho\right)  \right\vert
^{2}\left(  n_{4}-\left\vert C\left(  \rho\right)  \right\vert ^{2}\right)
\nonumber\\
&  +n_{12}\left\vert C\left(  \rho\right)  \right\vert ^{2}, \label{f16}%
\end{align}
To obtain Eq. (\ref{n4rho}) we have used the expressions for the coefficients
$n_{d}$ ($d=4,8,12,16$) in terms of eigenvalues of matrix $\left(
\rho\widetilde{\rho}\right)  $. For qubit one and qubit number $j$ in state
$\rho_{1j}$ if $C\left(  \rho_{1j}\right)  \geq0$, then $\tau_{1|j}\left(
\rho_{1j}\right)  =\left\vert C\left(  \rho_{1j}\right)  \right\vert $
otherwise $\tau_{1|2}\left(  \rho_{1j}\right)  =0$. Therefore, we may rewrite
Eq. (\ref{n4rho}) as%
\begin{equation}
n_{4}\left(  \rho_{1j}\right)  -\tau_{1|2}^{2}\left(  \rho_{1j}\right)
=\sqrt{4n_{8}\left(  \rho_{1j}\right)  +\chi_{1j}^{\pm}}, \label{2tanrho}%
\end{equation}
where $\chi_{1j}^{+}=8\left(  \sqrt{n_{16}\left(  \rho_{1j}\right)  }%
+\sqrt{f_{16}\left(  \rho_{1j}\right)  }\right)  $, and%
\begin{equation}
\chi_{1j}^{-}=\left(
\begin{array}
[c]{c}%
8\sqrt{n_{16}\left(  \rho_{1j}\right)  }-8\sqrt{f_{16}\left(  \rho
_{1j}\right)  }\\
+2n_{4}\left(  \rho_{1j}\right)  \left\vert C\left(  \rho_{1j}\right)
\right\vert ^{2}-\left\vert C\left(  \rho_{1j}\right)  \right\vert ^{4}%
\end{array}
\right)  .
\end{equation}
This is an important relation between coefficients $n_{d}\left(  \rho
_{1j}\right)  $ and two-tangle.


\paragraph*{\textbf{\textit{Two-tangles and one-tangle of an N-qubit pure
State:}}}

In this section, we consider the case where two-qubit state is a marginal
state of an $N-$qubit pure state. Our main objective is to find a relation
between two-tangles of all the qubit pairs with one tangle of the focus qubit.
A general $N-$qubit pure state in computational basis reads as%
\begin{equation}
\left\vert \Psi_{12...N}\right\rangle =\sum_{i_{1},i_{2},...,N}a_{i_{1}%
i_{2}...i_{N}}\left\vert i_{1}i_{2}...i_{N}\right\rangle ;\ i_{m}=0,1.
\label{psiN}%
\end{equation}
Here $a_{i_{1}i_{2}...i_{N}}$ are complex state coefficients and the indices
$i_{1}$, $i_{2},...,i_{N}$ refer to the state of qubits at locations $A_{1}$,
$A_{2}$,..., $A_{N}$, respectively. State of qubit pair $A_{1}A_{j}$ is
$\rho_{1j}=Tr_{2,...,j-1,j+1,...N}(\left\vert \Psi_{12...N}\right\rangle
\left\langle \Psi_{12...N}\right\vert )$ with matrix elements given by%
\begin{equation}
\left(  \rho_{1j}\right)  _{i_{1}i_{j}k_{1}k_{j}}=\sum_{I}a_{i_{1}i_{j}%
I}a_{k_{1}k_{j}I}^{\ast},
\end{equation}
where index $I=\left\{  i_{2}i_{3}...i_{j-1}i_{j+1}...i_{N}\right\}  $ with
associated value $I_{v}\equiv%
{\displaystyle\sum\limits_{\substack{m=2\\m\neq j}}^{N}}
2^{m-2}i_{m}$. Writing the characteristic polynomial for $\rho_{1j}%
\widetilde{\rho}_{1j}$ it is found that for qubit pair $A_{1}A_{j}$ in state
$\rho_{1j}$,%
\begin{equation}
n_{4}\left(  \rho_{1j}\right)  =2\sum_{I\leq J}\left\vert D_{1jIJ}%
+D_{1jJI}\right\vert ^{2}, \label{n4ro1j}%
\end{equation}
where $D_{1jIJ}=a_{0\left(  i_{j}=0\right)  I}a_{1\left(  i_{j}=1\right)
J}-a_{1\left(  i_{j}=0\right)  J}a_{0\left(  i_{j}=1\right)  I}$. The
simplified notation $I<J$ is being used to indicate that $I_{v}<J_{v}$. The
functions $\left(  D_{1jIJ}+D_{1jJI}\right)  $ are invariant with respect to
unitary transformations on the focus qubit and qubit $j$ and depending on the
value of $I$ and $J$ represent a sum of determinants of $2-$way, $3-$way,...,
$N-$way matrices of dimension $2$. One-tangle defined as $\tau_{1|2...N}%
=4\det\left(  \rho_{1}\right)  $ where $\rho_{1}=Tr_{A_{2}...A_{N}}(\left\vert
\Psi_{12...N}\right\rangle \left\langle \Psi_{12...N}\right\vert )$,
quantifies the entanglement of qubit $A_{1}$ with rest of the system. One can
verify that%
\begin{equation}
\tau_{1|2...N}=4\sum_{j=2}^{N}\left(  \sum_{I}\left\vert D_{1jII}\right\vert
^{2}+\sum_{I<J}\left\vert D_{1jIJ}\right\vert ^{2}\right)  . \label{1tanN}%
\end{equation}
By definition, $\tau_{1|j}^{2}\left(  \rho\right)  \leq4\sum_{I}\left\vert
\left(  D_{1jII}\right)  \right\vert ^{2}$, therefore%
\begin{equation}
\tau_{1|2...N}\geq\sum_{j=2}^{N}\tau_{1|j}^{2}\left(  \rho\right)  ,
\label{1tanosbo}%
\end{equation}
which is the inequality of ref. \cite{osbo06}. Comparing Eq. (\ref{n4ro1j})
and Eq. (\ref{1tanN}), we obtain%
\begin{equation}
\tau_{1|2...N}=\sum_{j=2}^{N}\left(  n_{4}\left(  \rho_{1j}\right)
-X_{1j}\right)  . \label{1tann4}%
\end{equation}
The sum of coherences, $\sum_{j=2}^{N}X_{1j}$, turns out to be zero for odd
values of $N$ and is equal to sum of unitary invariants of degree two for $N$ even.

The coefficient $n_{8}\left(  \rho_{1j}\right)  $, written in terms of state
coefficients reads as%
\begin{equation}%
\begin{array}
[c]{c}%
n_{8}\left(  \rho_{1j}\right)  =\\
\sum_{I,J,K,L}\left\vert
\begin{array}
[c]{c}%
\left(  D_{1jIJ}+D_{1jJI}\right)  \left(  D_{1jKL}+D_{1jLK}\right) \\
-\left(  D_{1jIL}+D_{1jLI}\right)  \left(  D_{1jKJ}+D_{1jJK}\right)
\end{array}
\right\vert ^{2}.
\end{array}
\label{n8ro1j}%
\end{equation}
While the coefficient $n_{4}\left(  \rho_{1j}\right)  $ is a sum of squares of
moduli of two-qubit invariants, the coefficient $n_{8}\left(  \rho
_{1j}\right)  $ is a sum of squares of three-qubit invariants.


\paragraph*{\textbf{\textit{Multipartite entanglement:}}}

Consider the special case where $\left\vert \Psi_{12...N}\right\rangle $ is a
product of two-qubit pure state and $N-2$ qubit pure state that is%
\begin{equation}
\left\vert \Psi_{12...N}\right\rangle =\sum_{i_{1},i_{j},I}a_{i_{1}i_{j}}%
b_{I}\left\vert i_{1}i_{j}I\right\rangle .
\end{equation}
In this case $D_{1jIJ}=D_{1jJI}=\left(  a_{00_{j}}a_{11_{j}}-a_{10_{j}%
}a_{01_{j}}\right)  b_{I}b_{J}$, therefore $n_{4}\left(  \rho_{1j}\right)
\neq0$ if qubits $A_{1}$ and $A_{j}$ are entangled, while $n_{8}\left(
\rho_{1j}\right)  =0$. This implies that to ensure that the state has
entanglement due to correlations beyond two-way correlations, the two-qubit
marginal states $\rho_{1j}$ of an $N-$qubit pure state must satisfy the
conditions, $tr\left(  \rho_{1j}\widetilde{\rho_{1j}}\right)  \neq0$, for
$j=2,...,N$, as well as%
\begin{equation}
\left(  tr\left(  \rho_{1j}\widetilde{\rho_{1j}}\right)  \right)
^{2}-tr\left(  \rho_{1j}\widetilde{\rho_{1j}}\right)  ^{2}\neq0.
\end{equation}


\paragraph*{\textbf{\textit{Three-tangles and the coefficient }}$n_{8}%
$\textbf{\textit{:}}}

Three tangle \cite{coff00} of a three-qubit pure state%
\begin{equation}
\left\vert \Psi_{123}\right\rangle =\sum\limits_{i_{1},i_{2},i_{3}}%
a_{i_{1}i_{2}i_{3}}\left\vert i_{1}i_{2}i_{3}\right\rangle ,\quad\left(
i_{m}=0,1\right)  , \label{3qstate}%
\end{equation}
is equal to four times the modulus of a unitary invariant polynomial of degree
four that is%
\begin{equation}
\tau_{1|2|3}=4\left\vert \left(  D_{1201}+D_{1210}\right)  ^{2}-4D_{1200}%
D_{1211}\right\vert .
\end{equation}
The entanglement measure $\tau_{1|2|3}\left(  \left\vert \Psi_{123}%
\right\rangle \right)  $ is extended to a mixed state $\rho_{123}$ of three
qubits via convex roof extension that is%
\begin{equation}
\tau_{1|2|3}\left(  \rho_{123}\right)  =\min_{\left\{  p_{i},\left\vert
\phi_{123}^{\left(  i\right)  }\right\rangle \right\}  }\sum\limits_{i}%
p_{i}\tau_{1|2|3}\left(  \left\vert \phi_{123}^{\left(  i\right)
}\right\rangle \right)  , \label{3tanmix1}%
\end{equation}
where minimization is taken over all complex decompositions $\left\{
p_{i},\left\vert \phi_{123}^{\left(  i\right)  }\right\rangle \right\}  $ of
$\rho_{123}$. Here $p_{i}$ is the probability of finding the normalized
three-qubit state $\left\vert \phi_{123}^{\left(  i\right)  }\right\rangle $
in the mixed state $\rho_{123}$.

A close look at $n_{8}\left(  \rho_{1j}\right)  $ ( Eq. (\ref{n8ro1j})) shows
it to be a function of three-tangles $\tau_{1|j|k}^{2}\left(  \rho
_{1jk}\right)  $, where $k\neq j$ varies from $2$ to $N$, along with
additional three-qubit invariants. In particular for $N=4$, a comparison of
$n_{8}\left(  \rho_{1j}\right)  $ with the upper bound on three-tangles of a
four-qubit pure state from ref. \cite{shar17} reveals that $n_{8}\left(
\rho_{1j}\right)  $ is related to three-tangles $\tau_{1|j|k}\left(
\rho_{1jk}\right)  $ through%
\begin{equation}
4n_{8}\left(  \rho_{1j}\right)  =\frac{1}{4}%
{\displaystyle\sum\limits_{k=2,k\neq j}^{4}}
\tau_{1|j|k}^{2}\left(  \rho_{1jk}\right)  +\delta_{1j},\label{4n812c4}%
\end{equation}
where $\delta_{1j}$ is a function of four-tangles. Details of the calculation
for four qubits are available in ref. \cite{arXiv2021v3}. The relation of
$n_{8}\left(  \rho_{1j}\right)  $ with three tangles of triples in an
$N-$qubit state can be obtained by successive applications of the method used
to obtain upper bound on three-tangle of a marginal state of four-qubit pure
state. We conjecture that if $\rho_{1j}$ is a two-qubit marginal state of
$N-$qubit pure state $\left\{  j=2,3,...,N\right\}  $ then the coefficient
$n_{8}\left(  \rho_{1j}\right)  $ (Eq. (\ref{n8ro1j}) can be written in the
form%
\begin{equation}
4n_{8}\left(  \rho_{1j}\right)  =\frac{1}{4}\sum_{k=2;k\neq j}^{N}\tau
_{1|j|k}^{2}\left(  \rho_{1jk}\right)  +\delta_{1j}.\label{4n81jcN}%
\end{equation}
A constraint on three-tangles is obtained by summing up degree eight
coefficients for all the pairs with $A_{1}$ as focus qubit, that is%
\begin{equation}
4%
{\displaystyle\sum\limits_{j=2}^{N}}
\left(  n_{8}\left(  \rho_{1j}\right)  -\frac{1}{4}\sum_{k=2;k\neq j}^{N}%
\tau_{1|j|k}^{2}\left(  \rho_{1jk}\right)  \right)  =%
{\displaystyle\sum\limits_{j=2}^{N}}
\delta_{1j}\label{4n8cN}%
\end{equation}
Monogamy constraint of Eq. (\ref{4n8cN}) does not allow a free distribution of
correlations between subsystems with $N>3$ and three-qubit subsystems.


\paragraph*{\textbf{\textit{Constraint on two-tangles and three tangles:}}}

Substituting $n_{8}\left(  \rho_{1j}\right)  $ from Eq. (\ref{4n81jcN}) into
Eq. (\ref{2tanrho}) written for $\rho_{1j}$ ($j=2,3,...N$), we obtain%
\begin{equation}%
\begin{array}
[c]{c}%
n_{4}\left(  \rho_{1j}\right)  -\tau_{1|j}^{2}\left(  \rho_{1j}\right)
=\sqrt{\frac{1}{4}\sum_{\substack{k=2\\k\neq j}}^{N}\tau_{1|j|k}^{2}\left(
\rho_{1jk}\right)  +\Delta_{1j}},\\
\Delta_{1j}=\delta_{1j}+\chi_{1j}^{\pm}\ ,
\end{array}
\label{n41jcN}%
\end{equation}
A constraint on two-tangles and three-tangles is obtained by taking the square
of Eq. (\ref{n41jcN}) and summing up over $j$ that is%
\begin{equation}%
\begin{array}
[c]{c}%
\sum_{j=2}^{N}\left(  n_{4}\left(  \rho_{1j}\right)  -\tau_{1|j}^{2}\left(
\rho_{1j}\right)  \right)  ^{2}-\frac{1}{2}\sum_{\substack{j,k=2\\j<k}%
}^{N}\tau_{1|j|k}^{2}\left(  \rho_{1jk}\right) \\
=\sum_{j=2}^{N}\Delta_{1j},
\end{array}
\label{n4cN}%
\end{equation}
where $\sum_{j=2}^{N}\Delta_{1j}$, represents residual correlations beyond
three-way correlations.


\paragraph*{\textbf{\textit{Constraint on tangles and }}$N$%
\textbf{\textit{-way correlations:}}}

A monogamy constraint on two-tangles, three-tangles and residual correlations
in an $N-$qubit state is obtained by substituting the expression for
$n_{4}\left(  \rho_{1j}\right)  $ from Eq. (\ref{2tanrho}) and $n_{8}\left(
\rho_{1j}\right)  $ from Eq. (\ref{4n81jcN}) in Eq. (\ref{1tann4}) that is%
\begin{align}
&  \tau_{1|23...N}+\sum_{j=4}^{N}X_{1j}-\sum_{j=2}^{N}\tau_{1|j}^{2}\left(
\rho_{1j}\right) \nonumber\\
&  =\sum_{j=2}^{N}\sqrt{\frac{1}{4}\sum_{k=2;k\neq j}^{N}\tau_{1|j|k}%
^{2}\left(  \rho_{1jk}\right)  +\Delta_{1j}}. \label{1tancN}%
\end{align}
Since for $N=3$, $4n_{8}\left(  \rho_{12}\right)  =4n_{8}\left(  \rho
_{13}\right)  =\frac{1}{4}\tau_{1|2|3}^{2}$, $\Delta_{1j}=0$, and $\sum
_{j=2}^{N}X_{1j}=0$ one verifies that the constraint of Eq. (\ref{1tancN}) is
equivalent to CKW inequality:%
\begin{equation}
\tau_{1|23}=\tau_{1|2}^{2}\left(  \rho_{12}\right)  +\tau_{1|3}^{2}\left(
\rho_{13}\right)  +\tau_{1|2|3}. \label{ckw}%
\end{equation}
In other words, with qubit $A_{1}$ as focus qubit the sum of two-tangles and
three-way correlations in $\left\vert \Psi_{123}\right\rangle $ is equal to
one tangle. Analogous relations are found by taking $A_{2}$ or $A_{3}$ as the
focus qubit. This also implies that if three-way correlations are maximal that
is $\tau_{1|2|3}=1=\tau_{1|23}$, then $\tau_{1|2}^{2}\left(  \rho_{12}\right)
=\tau_{1|3}^{2}\left(  \rho_{13}\right)  =0$.

Recalling that $n_{4}\left(  \rho_{1j}\right)  $ is a calculable quantity, for
$N=4$ the measures of two-way, three-way and four-way correlations satisfy the
condition (Eq. (\ref{n4cN})):%
\begin{equation}%
\begin{array}
[c]{c}%
\sum_{j=2}^{4}\left(  n_{4}\left(  \rho_{1j}\right)  -\tau_{1|j}^{2}\left(
\rho_{1j}\right)  \right)  ^{2}\\
-\frac{1}{2}\sum_{\substack{j,k=2\\j<k}}^{4}\tau_{1|j|k}^{2}\left(  \rho
_{1jk}\right)  =\sum_{j=2}^{4}\Delta_{1j}.
\end{array}
\label{n4c4}%
\end{equation}
An additional monogamy constraint on sub-system tangles of a four-qubit pure
state is obtained from Eq. (\ref{1tancN}) (details in ref. \cite{arXiv2021v3}
). These relations are found to be valid for all classes of four-qubit states
(\cite{vers02}).

From Eq. (\ref{1tancN}) one can also deduce that%
\begin{equation}%
\begin{array}
[c]{c}%
\tau_{1|23...N}-\sum_{j=2}^{N}\tau_{1|j}^{2}\left(  \rho_{1j}\right) \\
-\frac{1}{2}\left(  \sum_{j=2}^{N}\sum_{k=2,k\neq j}^{N}\tau_{1|j|k}%
^{2}\left(  \rho_{1jk}\right)  \right)  ^{\frac{1}{2}}\geq0\text{.}%
\end{array}
\label{1tanc2N}%
\end{equation}
The difference between one tangle and contributions from two-tangles and
three-tangles, represents the residual correlations beyond three-way
correlations present in an N-qubit pure state.


\paragraph*{\textbf{\textit{Concluding Remarks:}}}

The functional relations between tangles of a multiqubit composite system
reported in this letter,\ constrain the free distribution of entanglement
among the parties. Our main result is a set of constraints on one-tangle of a
focus qubit, two-tangles, three-tangles and residual correlations beyond
three-tangles, contained in Eqs. (\ref{4n81jcN} - \ref{1tancN}),
and Eq. (\ref{1tanc2N}). The residual correlations obtained by subtracting
two-tangles and three-tangles as in Eqs. (\ref{n4cN}) represent contributions
from all possible $N-$qubit entanglement modes. One tangle of a four-qubit
pure state satisfies the constraints obtained by substituting $N=4$ in Eqs.
(\ref{n4cN}) and (\ref{1tancN}) independent of the class \cite{vers02} to
which a given four-qubit state belongs. In particular, it has been shown in
ref. \cite{arXiv2021v3} that these constraints are satisfied by the set of
states $L_{a,ia,\left(  ia\right)  _{2}}$ that violate the entanglement
monogamy relation obtained by generalizing the CKW inequality.

This work reveals constraints on the sharing of entanglement at multiple
levels and offers insight into quantification of those features of quantum
correlations, which only emerge beyond the bipartite scenario. Any functional
relation between measures of entanglement in different subsets of parties is a
monogamy constraint because the free distribution of entanglement between
different parties is constrained by it. Monogamy constraints are useful to
investigate the interplay between the entanglement trade-off and frustration
phenomena in complex quantum systems \cite{giam11}. Our approach also paves
the way to understanding scaling of entanglement distribution as qubits are
added to obtain larger multiqubit quantum systems. Monogamy constraints also
help understand quantum to classical transition \cite{bran15}. If one part of
the composite system becomes highly entangled with its surroundings, its
entanglement with other part tends to zero as the quantum correlations between
two parts of the composite system leak into the environment \cite{entra}.
Consequently, components of the wave function are decoupled from a coherent
system and acquire phases from their immediate surroundings.

\textit{Acknowledgement: } We thank Capes, Brasil and UEL, PR Brasil, for
providing access to research journals. These authors have contributed to this work:

\dag shelly@uel.br (Associate Professor (retired ), Universidade estadual de Londrina),

\ddag nsharma@uel.br (Full Professor (retired ), Universidade estadual de Londrina).


\begin{thebibliography}{99}                                                                                               %


\bibitem {schr35}E. Schr\"{o}dinger, Discussion of probability relations
between separated systems, Proc. Cambridge Philos. Soc. 31, 555 (1935).

\bibitem {shor94}P. W. Shor, Algorithms for quantum computation: Discrete
logarithms and factoring, in Proceedings 35th Annual Symposium on Foundations
of Computer Science (IEEE, New York, 1994), pp. 124--134.


\bibitem {benn84}C. H. Bennett and G. Brassard, Quantum cryptography: Public
key distribution and coin tossing,, in Proceedings ofthe IEEE International
Conference on Computers, Systems and Signal Processing, Bangalore, India
(IEEE, New York, 1984), pp. 175--179.


\bibitem {cala12}P. Calabrese, J. Cardy, and E. Tonni, Entanglement negativity
in quantum field theory Phys. Rev. Lett. 109, 130502 (2012).


\bibitem {sahl15}S. Sahling, G. Remenyi, C. Paulsen, P. Monceau, V. Saligrama,
C. Marin, A. Revcolevschi, L. P. Regnault, S. Raymond and J. E. Lorenzo,
Nature Physics 11, 255--260 (2015).


\bibitem {lamb13}N. Lambert, Y. N. Chen, Y. C. Chen, C. M. Li, G. Y. Chen, and
F. Nori, Quantum biology, Nat. Phys. 9, 10 (2013).


\bibitem {xue02}Xue, P., Li, C. F. \& Guo, G. C. Conditional efficient
multiuser quantum cryptography network. Phys. Rev. A 65, 022317 (2002).


\bibitem {xue17}Xue, P., Wang, K. \& Wang, X. Efficient multiuser quantum
cryptography network based on entanglement. Sci Rep 7, 45928 (2017).


\bibitem {yang20}Yang, H., Xiao, M. Multi-user quantum private query. Quantum
Inf Process 19, 253 (2020).


\bibitem {louk10}Loukopoulos K., Browne D.E. Secure multiparty computation
with a dishonest majority via quantum means. Phys. Rev. A 81, 062336(2010).


\bibitem {ma08}Ma H., Chen B., Guo Z., Li H. Development of quantum network
based on multiparty quantum secret sharing. Can. J. Phys. 86, 1097--1101.
(2008).


\bibitem {horo09}R. Horodecki, P. Horodecki, M. Horodecki and K. Horodecki,
Rev. Mod. Phys. 81, 865 (2009).
.

\bibitem {walt13}M. Walter, B. Doran, D. Gross and M. Christandl, Science 340
(6137), 1205-1208 (2013).


\bibitem {hill97}S. Hill and W. K. Wootters, Phys. Rev. Lett. 78, 5022 (1997).

\bibitem {woot98}W. K. Wootters, Phys. Rev. Lett. 80, 2245 (1998).

\bibitem {coff00}V. Coffman, J. Kundu, and W. K. Wootters, Phys. Rev. A 61,
052306 (2000).


\bibitem {osbo06}T. J. Osborne and F. Verstraete, Phys. Rev. Lett. 96, 220503 (2006).

\bibitem {arXiv2021v3}S. S. Sharma and N. K. Sharma, arXiv:2002.00701v3
[quant-ph](2021). Supplementary article co-submitted to PRA.


\bibitem {regu14}Regula, B., Martino, S.D., Lee, S., Adesso, G.: Strong
monogamy conjecture for multiqubit entanglement: the four-qubit case. Phys.
Rev. Lett. 113, 110501 (2014)

\bibitem {regu16}B. Regula, A. Osterloh and G. Adesso, Phys. Rev. A 93, 052338 (2016)

\bibitem {karm16}Karmakar, S., Sen, A., Bhar, A., Sarkar, D.: Strong monogamy
conjecture in a four-qubit system. Phys. Rev. A 93, 012327 (2016)


\bibitem {shar18}S. S. Sharma and N. K. Sharma, Quantum Inf Process, Vol. 17,
183 (2018).


\bibitem {park18}D. Park, arXiv:1801.07846 [quant-ph].

\bibitem {chri18}Christopher Eltschka and Jens Siewert, Quantum 2, 64
(2018).$\allowbreak$ $\allowbreak$
.

\bibitem {terh04}B. M. Terhal, IBM Journal of Research and Development, 48 (1)
: 71- 78, (2004).


\bibitem {pawl10}M. Pawlowski.
Phys. Rev. A 82 032313 (2010).


\bibitem {gisi02}N. Gisin, G. Ribordy, W. Tittel, and H. Zbinden, Rev. Mod.
Phys., 74:145, (2002).


\bibitem {dur00}W. D\"{u}r, G. Vidal, and J. I. Cirac, PRA 62 062314 2000.


\bibitem {gior11}G. L. Giorgi, Phys. Rev. A, 84: 054301, (2011).


\bibitem {prab12}R. Prabhu, A. K. Pati, A. SenDe, and U. Sen, Phys. Rev. A,
85:040102(R), (2012).


\bibitem {ma11}X. Ma, B. Dakic, W. Naylor, A. Zeilinger and P. Walther, Nature
Phys 7, 399--405 (2011).


\bibitem {rao13}K. R. Rao, H. Katiyar, T. S. Mahesh, A. SenDe, U. Sen, and A.
Kumar, Phys. Rev. A, 88:022312, (2013).


\bibitem {lloy14}S. Lloyd and J. Preskill, J. High Energy Phys., 08, 126
(2014).


\bibitem {shar16}S. S. Sharma and N. K. Sharma, Quantum Inf Process, Vol. 15,
4973--4993 (2016).


\bibitem {shar17}S. S. Sharma and N. K. Sharma, Phys. Rev. A 95,
062311(2017).


\bibitem {vers02}F. Verstraete, J. Dehaene, B. De Moor, and H. Verschelde,
Phys. Rev. A 65, 052112 (2002).

\bibitem {giam11}S. M. Giampaolo, G. Gualdi, A. Monras, and F. Illuminati,
Phys. Rev. Lett. 107, 260602 (2011).

\bibitem {bran15}Brandao, F., Piani, M., and Horodecki, P. Generic emergence
of classical features in quantum Darwinism. Nat Commun 6, 7908 (2015).

\bibitem {entra}*Please check a simple circuit model in \cite{arXiv2021v3} to
illustrate this point.
\end{thebibliography}
\end{document}